\documentclass[12pt,thmsa]{article}
\usepackage{sw20lart}

\input{tcilatex}
\input tcilatex

\begin{document}

\author{BABUR M. MIRZA \\
Department of Mathematics, Quaid-i-Azam University,\\
\ Islamabad 45320, Pakistan\\
E-mail: bmmirza2002@yahoo.com}
\title{Symmetry Reduction of Lane-Emden Equation for Polytropes}
\date{October 16, 2005}
\maketitle

\begin{abstract}
We describe an ansatz for symmetry reduction of the Lane-Emden equation for
an arbitrary polytropic index $n$, admitting only one symmetry generator.
For the reduced first order differential equation it is found that standard
reduction procedure do not admit any non-trivial Lie point symmetry. However
some special solutions for the differential equation are obtained.
\end{abstract}

The simplest model of a gravitationally bound spherically symmetric
polytrope with a power law distribution function is determined by the
Lane-Emden equation\cite{[1],[2]} 
\begin{equation}
\frac 1{r^2}\frac d{dr}(r^2\frac{d\psi (r)}{dr})=-\psi (r)^n
\end{equation}
where $\psi (r)$ is the dimensionless gravitational potential, $r$ is the
dimensionless radius of the sphere, and $n$ is the polytropic index. For the
second order differential equation the symmetry generator has the form\cite
{[3]}

\begin{equation}
\mathbf{X}=\xi (r,\psi )\frac \partial {\partial r}+\eta (r,\psi )\frac
\partial {\partial \psi }+\eta ^{\prime }(r,\psi )\frac \partial {\partial
\psi ^{\prime }}
\end{equation}
Here the prime denotes differentiation with respect to the independent
variable $r$. For the differential equation the symmetry condition is given
by

\begin{equation}
\mathbf{X}\omega (r,\psi ,\psi ^{\prime })=\eta ^{\prime \prime }(r,\psi
),\quad \quad \func{mod}\psi ^{\prime \prime }=\omega (r,\psi ,\psi ^{\prime
}).
\end{equation}
This gives an identity in the dependent and independent variables $(r,\psi )$%
. From (1) we have

\begin{equation}
\omega (r,\psi ,\psi ^{\prime })=-\psi (r)^n-\frac 2r\psi ^{\prime }.
\end{equation}
The infinitesimals $(\xi ,\eta )$ are now determined by the identity

\[
(\psi ^n+\frac 2r\psi ^{\prime })[\eta ,_\psi -2\xi ,_r-3\psi ^{\prime }\xi
,_\psi ]+\frac 2{r^2}\psi ^{\prime }\xi -n\psi ^{n-1}\eta -\frac 2r[\eta
,_r+\psi ^{\prime }(\eta ,_\psi -\xi ,_r) 
\]
\begin{equation}
-(\psi ^{\prime })^2\xi ,_\psi ]=\eta ,_{rr}+\psi ^{\prime }[2\eta ,_{r\psi
}-\xi ,_{rr}]+(\psi ^{\prime })^2[\eta ,_{\psi \psi }-2\xi ,_{r\psi }]-(\psi
^{\prime })^3\xi ,_{\psi \psi }
\end{equation}
We can determine the infinitesimals $\xi $ and $\eta $ by an iterative
procedure. First equating coefficients of $(\psi ^{\prime })^3$ and $(\psi
^{\prime })^2$ in (5), we find that $\xi \neq \xi (\psi )$ and $\eta ,_{\psi
\psi }=0$. Therefore we assume that $\xi =\alpha (r)$ and $\eta =\beta
(r)\psi +\gamma (r)$. The symmetry condition then reduces to

\[
(\psi ^n+\frac 2r\psi ^{\prime })[\beta -2\alpha ^{\prime }]+\frac
2{r^2}\psi ^{\prime }\alpha -n\psi ^{n-1}(\beta \psi +\gamma )-\frac
2r[\beta ^{\prime }\psi +\gamma ^{\prime }+\psi ^{\prime }(\beta -\alpha
^{\prime })] 
\]

\begin{equation}
=\beta ^{\prime \prime }\psi +\gamma ^{\prime \prime }+\psi ^{\prime
}[2\beta ^{\prime }-\alpha ^{\prime \prime }]
\end{equation}
Equating the coefficients of $\psi ^{\prime }$ and $\psi ^n$ we obtain the
following two conditions

\begin{equation}
2\alpha -2r\alpha ^{\prime }-[2\beta ^{\prime }-\alpha ^{\prime \prime
}]r^2=0,\quad [\beta -2\alpha ^{\prime }]-n\beta =0
\end{equation}
The coefficients of $r^2$ equated to zero leave for the first two terms in
the first condition $\alpha $ $=kr$ where $k$ does not depend on $r$.The
second equation gives $\beta =2\alpha ^{\prime }/(1-n)$. Substitution back
into the determining equation yields

\begin{equation}
n\gamma \psi ^{n-1}+\frac 2r\gamma ^{\prime }+\gamma ^{\prime \prime }\equiv
0
\end{equation}
Comparing coefficients of $\psi ^{n-1}$ we find that $\gamma =0$. Thus the
infinitesimals are

\begin{equation}
(\xi ,\eta )=(kr,\frac{-2k}{n-1}\psi ),\quad n\neq 1.
\end{equation}
We have here a one-parameter group $G_1$ of scaling transformation with a
parameter $k$. Thus reduction in the order of the equation by one is possible
\cite{[4]}. For convenience we take $k$ to be unity.

Let $(t,s)$ be the canonical variables for the differential equation (1)
with infinitesimals (9). Then

\begin{equation}
s=\int \frac{dx}\xi ,\quad t=\int (\xi dy-\eta dx);
\end{equation}
gives

\begin{equation}
s=\ln x,\quad t=y\exp (\frac 2{n-1});\quad n\neq 1.
\end{equation}
We have for $s=s(t)$ and $s^{\prime }=ds/dt$:

\begin{equation}
dr=d(e^s)=e^sds,\quad d\psi =d(te^{2s/(1-n)})=e^{2s/(1-n)}dt+\frac{2t}{1-n}%
e^{2s/(1-n)}ds.
\end{equation}
which implies that

\begin{equation}
\psi ^{\prime }=\frac{d\psi }{dr}=\frac{e^{(\frac{1+n}{1-n})s}}{s^{\prime }}[%
1+\frac 2{1-n}ts^{\prime }],
\end{equation}
and

\begin{equation}
\psi ^{\prime \prime }=\frac d{dt}(\frac{d\psi }{dr})=\frac{e^{(\frac{2n}{1-n%
})s}}{(s^{\prime })^3}[s^{\prime }\{\frac{2(ts^{\prime \prime }+s^{\prime })%
}{1-n}+\frac{1+n}{1-n}(1+\frac{2ts^{\prime }}{1-n})s^{\prime }\}-s^{\prime
\prime }(1+\frac{2ts^{\prime }}{1-n})]
\end{equation}
With these substitutions we obtain from equation (1) the following reduced
form of the Lane-Emden equation

\begin{equation}
u^{\prime }(t)=\frac{n-5}{n-1}u^2+\{\frac{2(3-n)}{(n-1)^2}t+t^n\}u^3,\quad
n\neq 1.
\end{equation}
Where $u(t)=s^{\prime }(t)$ and $u^{\prime }(t)=s^{\prime \prime }(t)$. This
highly nonlinear first order differential equation has solutions for $n=0$
and $5$. These correspond to the well known solutions for $\psi $ namely $%
-(r^2/6)+(1/6x)$ and $(1+r^2/3)^{-1/2}$ respectively. The only other known
solution to (15) is the spherical Bessel function corresponding to $n=1$\cite
{[5],[6]}. To investigate the solution for other permissible values of $n$
we first transform the variable $u(t)$ as $-1/y(t)$. This substitution
weakens the nonlinearity of the equation as

\begin{equation}
y^{\prime }(t)=a-\frac{bt+t^n}{y(t)},\quad n\neq 1,
\end{equation}
where $a=(n-5)/(n-1)$ and $b=2(3-n)/(n-1)^2$. For the infinitesimals $%
(\varsigma ,\phi )$The symmetry condition for equation (16) gives

\begin{equation}
\phi ,_{t}+(\phi ,_{y}-\varsigma ,_{t})(a-\frac{bt+t^{n}}{y(t)})-\varsigma
,_{y}(a-\frac{bt+t^{n}}{y(t)})^{2}=\varsigma (-\frac{b+nt^{n-1}}{y})+\phi (%
\frac{bt+t^{n}}{y^{2}})
\end{equation}
For the infinitesimals of this equation we find that no non-trivial solution
to (16) exists for the above ansatz.

However one immediate consequence of the infinitesimals in (9) is an
invariant particular solution for the Lane-Emden equation. We first have

\begin{equation}
\psi ^{\prime }(r)=\frac \eta \xi =\frac{-2}{n-1}(\frac \psi r),\quad \psi
^{\prime \prime }(r)=\frac{2(n+1)}{(n-1)^2}\frac \psi {r^2}
\end{equation}
Substitution into equation (1) we obtain the particular solution in an
implicit form

\begin{equation}
\psi ^{n}=\frac{2(n-3)}{(n-1)^{2}}\frac{\psi }{r^{2}},\quad n\neq 1.
\end{equation}
For $n$ less than one this solution is bounded for all values of $r$. For $n$
$\succ 1$ the solution has a singularity at $r=0$, which becomes stronger as 
$n$ increases.


\begin{thebibliography}{9}
\bibitem{[1]}  Liu F K, Polytropic gas spheres: An approximate analytic
solution of the Lane-Emden equation, \textit{MNRAS} \textbf{281} (1996),
1197.

\bibitem{[2]}  Honda M and Honda Y S, On Exact Polytropic Equilibria of
Self-Gravitating Gaseous and Radiative Systems: Their Application to
Molecular Cloud Condensation, \textit{MNRAS} \textbf{341} (2003), 164.

\bibitem{[3]}  Stephani H, Differential Equations: Thier Solution using
Symmetries, University Press, Cambridge, 1989, p28.

\bibitem{[4]}  Bluman G W and Kumei S, Symmetries and Differential
Equations. Applied Mathematical Sciences 81, Springer-Verlag, New York,
1989, p110.

\bibitem{[5]}  Rose W K, Advanced Stellar Astrophysics. University Press,
Cambridge, 1998, p447.

\bibitem{[6]}  McKee C F and Holliman J H II, Multi--Pressure Polytropes as
Models for the Structure and Stability of Molecular Clouds. I. Theory. 
\textit{ApJ.} \textbf{522} (1999), 313. 
\end{thebibliography}
\end{document}